\documentclass[11pt]{article}
\usepackage[centertags]{amsmath}
\usepackage[square, comma, sort&compress,numbers]{natbib}
\usepackage{array,multirow}
\numberwithin{equation}{section}
\usepackage{amssymb}
\usepackage{graphicx}
\usepackage{float}
\usepackage{color}
\usepackage{epsfig}
\usepackage{tikz,pgf}
\usetikzlibrary{shapes}
\usetikzlibrary{calc}
\usetikzlibrary{decorations.pathmorphing}
\usetikzlibrary{decorations.pathreplacing,shapes.misc}
\usetikzlibrary{positioning}
\usetikzlibrary{arrows}
\usetikzlibrary{decorations.markings}
\tikzset{snake it/.style={decorate, decoration=snake}}

\def\be{\begin{equation}}
\def\ee{\end{equation}}
\def\ba{\begin{array}}
\def\ea{\end{array}}
\newcommand{\bref}[1]{\textbf{\ref{#1}}}

\def\1{\tilde{1}}
\def\2{\tilde{2}}
\def\3{\tilde{3}}
\def\D{\Delta}
\def\td{\tilde\Delta}
\def\ep{\epsilon}
\def\ra{\rightarrow}
\def\ept{\tilde\epsilon}

\def\be{\begin{equation}}
\def\ee{\end{equation}}
\def\ba{\begin{array}}
\def\ea{\end{array}}

\def\infc{c \rightarrow  \infty}

\usepackage{jheppub}

\makeatletter
\def\@fpheader{\vspace{-.1cm}}
\makeatother

\title{Geodesic description of Heavy-Light Virasoro blocks}

\author[b,c,d]{Vladimir\ Belavin,}

\author[a]{Roman\ Geiko}

\affiliation[a]{National Research University Higher School of Economics,\\
Usacheva str. 6, 119048 Moscow, Russia}

\affiliation[b]{I.E. Tamm Department of Theoretical Physics, \\P.N. Lebedev Physical
Institute,\\ Leninsky ave. 53, 119991 Moscow, Russia}
\affiliation[c]{Moscow Institute of Physics and Technology, \\
Dolgoprudnyi, 141700 Moscow region, Russia}
\affiliation[d]{Department of Quantum Physics, \\
Institute for Information Transmission Problems, \\
Bolshoy Karetny per. 19, 127994 Moscow, Russia}

\emailAdd{rgeyko@hse.ru}
\emailAdd{belavin@lpi.ru}
\abstract{We continue to investigate the dual description of the Virasoro conformal blocks
arising in the framework of the classical limit of the AdS$_3$/CFT$_2$ correspondence.
To give such an interpretation in previous studies, certain restrictions were necessary.
Our goal here is to consider a more general situation available through the worldline
approximation to the dual AdS gravity. Namely, we are interested in computing the spherical
conformal blocks without the previously imposed restrictions on the conformal dimensions of
the internal channels. The duality is realised as an equality of the so-called heavy-light
limit of the $n$-point conformal block and the action of $n{-}2$ particles propagating in
some AdS-like background with either a conical singularity or a BTZ black hole.
We describe a procedure that allows relaxing the constraint on the intermediate channels.
To obtain an explicit expression for the conformal block on the CFT side, we use a recently
proposed recursion procedure and find full agreement between the results of the boundary
and bulk computations.
}

\keywords{Conformal field theory, Virasoro algebra, AdS/CFT duality}
\arxivnumber{}
\notoctrue{}

\begin{document}

\maketitle
\flushbottom

\section{Introduction}

Computing correlators in conformal field theory (CFT) requires knowing the
conformal blocks \cite{Belavin:1984vu}. They define the contributions of
particular sets of primary fields and their descendants to the intermediate
channels when the operator product expansion (insertion of a complete set of states) is applied.
Although symmetry algebras in principle fix the coefficients of the series expansions for these functions,
conformal blocks are known explicitly only in particular cases.
The AdS/CFT duality reveals an interesting interpretation of conformal blocks \cite{Fitzpatrick:2014vua}. Basically,
this interpretation is available in a semiclassical treatment (but see, e.g. \cite{Fitzpatrick:2015dlt,Aharony:2016dwx}).
Nevertheless, despite a certain progress in understanding the dual description of conformal blocks, there are still some open questions, one of which we address here.

We focus on AdS$_3$/CFT$_2$ in the case of a pure Virasoro symmetry. The main issue is to define
bulk degrees of freedom appropriate to the observables of the boundary CFT.
According to the Brown--Henneaux relation $c\sim 1/G_{N}$,
we are dealing with the large central charge CFT limit and the classical gravity in the bulk \cite{Brown:1986nw}.
The large central charge behaviour of the conformal blocks was first considered
in the context of their analytic properties \cite{Zamolodchikov:1987ie}.
It was noticed that the classical limit of the conformal block has
an interesting exponentiation property: it becomes an exponent of a function
depending on appropriately rescaled conformal dimensions (called classical).
This function is known as the {\it classical conformal block}.
To be interpreted as a classical action, a certain linearisation procedure \cite{Fitzpatrick:2014vua,Hijano:2015rla}
must be applied to the classical block. The obtained object is called a heavy-light block.
An important statement from the previous studies~\cite{Hartman:2013mia,Fitzpatrick:2015zha,Hijano:2015qja,Caputa:2014eta,Caputa:2015waa,Banerjee:2016qca,Chen:2016dfb,Hulik:2016ifr,Fitzpatrick:2016mjq,Alkalaev:2016rjl}
implies that the heavy-light block equals to the classical action up to some easily tractable terms.
To verify this statement, some methods of calculating the CFT counterpart of the duality were developed \cite{Bhatta:2016hpz,Besken:2016ooo,Alkalaev:2016fok}.
In this paper, we discuss the CFT on a sphere, but the duality can also be extended to conformal blocks on higher-genus surfaces \cite{Krasnov:2000zq,Kraus:2016nwo,Alkalaev:2016ptm,AlkBel4}.

We consider the AdS/CFT correspondence in the example of the 5-point heavy-light block.
The dual description \cite{Alkalaev:2015wia,Alkalaev:2015fbw,Alkalaev:2015lca} for this object is as following:
two {\it heavy} fields form the BTZ or conical singularity \cite{Banados:1992wn},
and three remaining {\it light} fields correspond to three particles propagating in this background (see Fig. \bref{Config}).
Such an interpretation prescribes that two dimensions of the heavy fields are equal.
Calculating the minimal action of the dual configuration involves solving a variation problem.
The solution can be found exactly when one of light external particles is absent
and the configuration hence corresponds to a 4-point heavy-light block.
The problem can then be solved using perturbation theory with the third particle mass regarded as a small parameter.
But one additional constraint on the internal dimensions was imposed by this method.
This approach led to the configuration with equal dimensions of the internal fields.
This point was not obvious, because there are no physical reasons for this constraint.
Hence, there is a question about principal possibility to solve this problem with unequal internal dimensions.

In this note, we compute the action corresponding to the 5-point heavy-light block.
We use an approach that allows analysing the bulk configuration perturbatively
without the previously imposed restriction on the internal channels.
We compare the obtained action with the boundary CFT calculation and find agreement.
The CFT side of the problem requires an effective method for computing conformal block coefficients.
The most convenient method for our purpose is $c$-recursion.
It was first developed for the 4-point conformal block by Zamolodchikov \cite{Zamolodchikov:1985ie}
in the form of the pole decomposition over the $c$-plane with the residues proportional to conformal blocks with shifted parameters.
This property leads to the recursion procedure of calculating the coefficients of conformal blocks level by level.
This idea was recently extended to $n$-point spherical and toric conformal blocks \cite{Cho:2017oxl}. We apply these results here to compute the 5-point block.

\section{Conformal side of conformal blocks}

We are interested in the conformal blocks of the correlation function with two heavy asymptotic states. Using projective invariance to fix $z_5 = \infty$, $z_4 =1$, and $z_1=0$
we obtain the block  depending on two modules $z_2$ and $z_3$.
The corresponding diagram is depicted on the Fig. \bref{5block}.
\begin{figure}[H]
\begin{center}
\begin{tikzpicture}\label{5block}

\draw [line width=1pt] (30,0) -- (38,0);
\draw [line width=1pt] (32,0) -- (32,2);
\draw [line width=1pt] (34,0) -- (34,2);
\draw [line width=1pt] (36,0) -- (36,2);


\draw (29.3,-0) node {$0, \Delta_1$};
\draw (32,2.5) node {$z_2, \Delta_{2}$};
\draw (34,2.5) node {$z_{3}, \Delta_{3}$};
\draw (36,2.5) node  {$1, \Delta_{h}$};
\draw (38.8,0) node {$\infty, \Delta_h$};

\draw (33,-0.5) node {$\widetilde{\Delta}_1$};
\draw (35,-0.5) node {$\widetilde{\Delta}_2$};


\fill  (32,0) circle (0.8mm);

\fill  (34,0) circle (0.8mm);

\fill  (36,0) circle (0.8mm);

\end{tikzpicture}
\end{center}
\caption{$5$-point conformal block.}
\label{5block}
\end{figure}
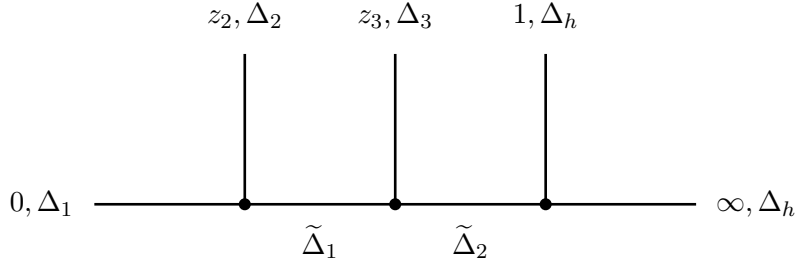
Introducing new variables $q_2 = z_3$ and  $q_1 = z_2/z_3$ and fixing intermediate channels, the conformal block
\be
\label{coef}
F(\Delta_{1,2,3,4,5}, \widetilde \Delta_{1,2}|q_1, q_2) = \sum_{r,s = 0}^\infty F_{r,s}\, q_1^r \, q_2^s\;,
\ee
with the expansion coefficients:
\begin{multline}
F_{r,s}=\sum_{r=|k|=|l|=0}^\infty \sum_{s=|m|=|n|=0}^\infty \mathcal{M}^{-1}_{k,l}(\td_1)\mathcal{M}^{-1}_{m,n}(\td_2)\langle
 \Delta_1| \mathcal{O}_2 \mathcal{L}_{-k} |\widetilde \Delta_1\rangle\ \times\\
  \langle \widetilde \Delta_1 | \mathcal{L}_{l} \mathcal{O}_3 \mathcal{L}_{-m} | \widetilde \Delta_2 \rangle \, \langle \widetilde \Delta_2| \mathcal{L}_{n} \mathcal{O}_h| \Delta_h\rangle\;.
\end{multline}
Here $\mathcal{M}_{m,n}(\td_{1,2})$ are Gram matrices elements for the corresponding Virasoro states, $\mathcal{L}_{-n}=L_{-k}^{i_k}...L_{-1}^{i_1}$ are elements of the universal enveloping of Virasoro algebra with $|n|=i_1+...+ki_k$ and $\mathcal{L}_{m}=\mathcal{L}_{-m}^\dagger$.
\subsection{Recursive representation}
 We provide here the recursive representation of the 5-point conformal block. The most efficient way of computing the coefficients of the block is so-called q-recursion (see \cite{Beccaria:2016vxq}), but for our purposes it is more convenient to use c-recursion because it is more suitable  for comparison with the bulk computation.

Let us shortly review the recursive representation of the 4-point block. Calculations based on the definition quickly becomes tedious  with the increasing of the level.
In \cite{Zamolodchikov:1985ie} the method based on the analysis of the analytical properties of 4-point blocks has been proposed. Following \cite{Kac:1984mq} it was noticed that conformal blocks, considered as function of $c$,  have only simple poles in special points:
\be\label{degc}
\ba{l}
c_{r,s}(\td)=13+6(b_{r,s}^2(\td)+b_{r,s}^{-2}(\td)),\\
 b_{r,s}^2(\tilde\Delta)=\frac{1}{1-r^2}\bigg(2\tilde\Delta+rs-1+\sqrt{(r-s)^2+4(rs-1)\tilde\Delta+4\tilde\Delta^2}\bigg)\;,
 \ea
\ee
where $\Delta$ is the intermediate dimension. These poles are related to degenerate values of the intermediate conformal dimension. If the fusion rules for the external fields are satisfied the  conformal block is regular. Consequently, the residues in points $c=c_{r,s}$ must contain so-called fusion polynomials which have zeros cancelling zeros of the denominator and can be derived from this requirement.

This idea can be generalized on an arbitrary number of intermediate channels \cite{Cho:2017oxl}. Here we provide necessary formulas for computing the 5-point conformal block.
The conformal block has poles in $c=c_{r,s}(\td_1)$ and $c=c_{r,s}(\td_2)$ \eqref{degc}
related to degenerate values in the first and in the second intermediate channels
\be
\label{Recursion}
F_{k,m}=G_{k,m}+\sum_{r\ge2 ,s\ge 1 }\frac{V_{r,s}^{(1)}}{c-c_{r,s}(\td_1)}+\sum_{r\ge2 ,s\ge 1}\frac{V_{r,s}^{(2)}}{c-c_{r,s}(\td_2)}\;,
\ee
where $V_{r,s}^{(1)}$ and $V_{r,s}^{(2)}$ are the corresponding residues and $G_{k,m}$ gives the asymptotics of $F_{k,m}$ as $\infc$. The residues are given explicitly by
\be
V_{r,s}^{(1)}=-\frac{\partial c_{r,s}(\td_1)}{\partial \td_1} A^{c_{rs}(\td_1)}_{rs}P^{rs}_{c_{rs}(\td_1)}\begin{bmatrix}\D_1\\ \D_2\end{bmatrix}P^{rs}_{c_{rs}(\td_1)}\begin{bmatrix}\td_2\\ \D_3\end{bmatrix}F_{k-rs,m}(\td_1\rightarrow \td_1+rs,c\rightarrow c_{rs}(\td_1))\;,\\
\ee
\be
V_{r,s}^{(2)}=-\frac{\partial c_{r,s}(\td_2)}{\partial \td_2}A^{c_{rs}(\td_2)}_{rs}P^{rs}_{c_{rs}(\td_2)}\begin{bmatrix}\D_4\\ \D_5\end{bmatrix}P^{rs}_{c_{rs}(\td_2)}\begin{bmatrix} \D_3\\ \td_1\end{bmatrix}F_{k,m-rs}(\td_2\rightarrow \td_2+rs,c\rightarrow c_{rs}(\td_2))\;,
\ee
where $A^{c_{rs}(\D)}_{rs}$ are the norms of the degenerated vectors:
\be
A^{c_{rs}(\D)}_{rs}=\frac{1}{2}\prod_{m=1-r}^r\prod_{n=1-s}^s(mb_{rs}(\D)+nb_{rs}(\D)^{-1})^{-1}, ~~(m,n)\neq(0,0),(r,s)\;,
\ee
and $P^{rs}_{c_{rs}(\D)}\begin{bmatrix} \D_i\\ \D_j \end{bmatrix}$ are the fusion polynomials:
\begin{multline}
P^{rs}_{c_{rs}(\D)}\begin{bmatrix} \D_i\\ \D_j\end{bmatrix}
=
\prod_{p=1-r ~{\rm step}~2}^{r-1}\, \prod_{q=1-s~{\rm step}~2}^{s-1}{\lambda^i_{rs}(\D)+\lambda^j_{rs}(\D)+pb_{rs}(\D)+qb_{rs}(\D)^{-1}\over2}\times\\
{\lambda^i_{rs}(\D)-\lambda^j_{rs}(\D)+pb_{rs}(\D)+qb_{rs}(\D)^{-1}\over2}\;,
\end{multline}
where
\be
\lambda^i_{rs}(\D)=\frac{\sqrt{(b_{rs}(\D)+b_{rs}(\D){-1})^2-4\D}}{4}\;.
\ee
The last ingredient is $G_{k,m}$ \cite{Alkalaev:2015fbw}:
\be
\label{coef2}
G_{k,m} =\frac{(\widetilde \Delta_1 + \Delta_2 -\Delta_1)_k \,(\widetilde \Delta_2 + \Delta_4 -\Delta_5)_m}{ (2\widetilde \Delta_1)_k \,(2\widetilde \Delta_2)_m}\,  \tau_{k,m}\;,
\ee
where the function $\tau_{k,m}$ is given by
\be
\label{tau2}
\tau_{k,m} = \sum_{p = 0}^{\min[k,m]} (-1)^p\frac{(-2\widetilde \Delta_2 -m+1)_{p}
(\widetilde \Delta_2+\Delta_3 - \widetilde \Delta_1)_{m-p}(\widetilde \Delta_1 + \Delta_3 -\widetilde \Delta_2+p-m)_{k-p}}{p!(k-p)!(m-p)!} \;.
\ee
These formulas provide an efficient way to compute the conformal block coefficients.

\subsection{Classical block}
Exponentiation of the conformal block in the semiclassical limit when all dimensions are heavy was firstly observed in \cite{Belavin:1984vu,Zamolodchikov:1987ie} and has natural physical interpretation in the framework of  Liouville field theory consideration. The deep relation between AdS$_3$ physics and Liouville theory explains our interest to the  classical blocks (see, e.g. \cite{Verlinde:1989ua,Krasnov:2001ui,Verlinde:2015qfa}). Our study of conformal blocks in this context is motivated by the goal to better understand this connection.

From the quantum block we obtain the classical block taking the $c\rightarrow \infty$ limit:
\begin{multline}
f_{cl}(\ep_{1,2,3},\ep_h, \ept_{1,2}|q_1,q_2)=\\
-\lim_{\substack{ c\to \infty}} \frac{6}{c}\log F(\D_{1,2,3} \ra  \frac{c}{6} \ep_{1,2,3},\D_h\ra \frac{c}{6}\ep_h,\td_{1,2}\ra \frac{c}{6}\ept_{1,2},c|q_1,q_2)\;.
\end{multline}
Let us notice that the existence of this limit imposes non-trivial constrains on the coefficients of the conformal blocks.

\subsection{Heavy-light block}
Here we define and calculate the heavy-light block. Firstly, we notice that the introduced notation can be confusing because the dimensions of the fields are usually classified according their central charge scaling properties. Now we introduce a smallness parameter $\delta$, which corresponds to the weakness of the gravity interactions between two particles. We call the first coefficient of the expansion of the classical block in $\delta$  the heavy-light block. Namely, we change
\be
\ep_{1,2,3}\rightarrow \delta \ep_{1,2,3}, \quad \ept_{1,2}\rightarrow \delta \ept_{1,2}, \quad \ep_{h}\rightarrow\ep_{h}\;,
\ee
and perform an expansion of the classical block in this parameter:
\be
f_{cl}=\delta f_{hl}+O(\delta^2)\;.
\ee
This procedure allows to take into account the fact that two fields are ``heavy'' in the sense of the gravitational  interaction. In other words, this takes into account the fact that two fields form the background and all others are ``light'' so that back-reaction is neglected.

In order to compare with the bulk computation we perform the following substitutions
\be
\ep_h=\frac{1-\alpha^2}{4},\quad\quad\ept_2=\ept_1+\mu \ep_3,\quad\quad \ep_2=\ep_1\;.
\ee
This yields  the following expansion
\be\label{f-hl-expan}
f_{hl}(\ep_1,\ep_1,\nu,\alpha, \ept_1, t|q_1,q_2)=-\sum_{m=0}\sum_{n=0}^{m}g_{mn}(\ep_1,\ept_1)\ep_3^m \mu^n\;.
\ee
Below the explicit results for the first three coefficients are presented:
\be
\ba{l}\label{g00}
g_{00}=\epsilon _1 \frac{q_1^2 q_2^2}{12}  \big(1 - \alpha ^2 \big)+\tilde{\epsilon }_1\big(\frac{1}{2} q_1 q_2 +\frac{5}{24} q_1^2
q_2^2 -\frac{1}{48} \alpha ^2 q_1^2 q_2^2 \big)\;,
\ea
\ee
\be
\ba{l}\label{g10}
g_{10}=\frac{q_1}{2}+\frac{q_2}{2}+\frac{q_1^2}{8}-\frac{q_1 q_2}{4}+\frac{7
q_2^2}{24}+\frac{q_1^3}{24}+\frac{q_1^2 q_2}{8} -\frac{q_1 q_2^2}{24} +\frac{5 q_2^3}{24}+\frac{q_1^4}{64}+\frac{q_1^3 q_2}{16} -\frac{11q_1^2 q_2^2}{96} -\frac{q_1 q_2^3}{48} +\frac{469 q_2^4}{2880}\\
+\alpha ^2\big(-\frac{q_2^2}{6}  +\frac{ q_1 q_2^2}{6} -\frac{q_1^2
q_2^2}{24}  -\frac{q_2^3}{6}  +\frac{ q_1 q_2^3}{12} -\frac{11}{72}
 q_2^4\big)+\frac{1}{180} \alpha ^4 q_2^4\;,
\ea
\ee
\be
\ba{l}\label{g11}
g_{11}=-\frac{q_1}{2}+\frac{q_2}{2}
-\frac{q_1^2}{8}+\frac{q_1 q_2}{4}+\frac{5
q_2^2}{24}
-\frac{q_1^3}{24}-\frac{q_1^2 q_2}{8} +\frac{q_1 q_2^2}{24} +\frac{q_2^3}{8}
-\frac{q_1^4}{64}-\frac{ q_1^3 q_2}{16}+\frac{11q_1^2 q_2^2}{96} +\frac{ q_1 q_2^3}{48}+\frac{251 q_2^4}{2880}\\
+\alpha^2 \big(-\frac{q_2^2}{12}+\frac{q_1 q_2^2}{12} -\frac{q_1^2 q_2^2}{48} -\frac{q_2^3}{12} +\frac{q_1 q_2^3}{24} -\frac{11
q_2^4}{144}\big)+\frac{7 \alpha ^4 q_2^4}{1440}\;.
\ea
\ee
Let us notice that the heavy-light block depends  linearly on the  light dimensions. This observation allows to think of  it as describing a solution of some classical problem. In the next section we will get an appropriate action.

\section{Bulk treatment}
Here we consider a classical gravity of probe particles in the AdS background with an angle defect. We provide necessary formulas in a very brief form (refer to \cite{Hijano:2015rla, Alkalaev:2015wia} for more details). We have a triple of coordinates $(t,\rho,\phi)$ and the interval is
\be
\label{metric}
ds^2  = \frac{\alpha^2}{\cos^2 \rho}\Big( - dt^2 +\sin^2\rho d\phi^2 +\frac{1}{\alpha^2} d\rho^2 \Big)\;,
\ee
where $\alpha$ corresponds to the conical defect. One can find the geodesic length for one particle in this background:
\be
S = \ln \frac{ \sqrt{\eta}}{\sqrt{1+\eta} +  \sqrt{1 - s^2 \eta}}\,\Bigg|_{\eta^{'}}^{\eta^{''}}\;,
\ee
with $\eta'$ and $\eta''$ being initial and final radial coordinates of the geodesic. Angular coordinate $\phi$ is cyclic and, as a consequence, we have conserved angular momenta. Here we have introduced the following notation
\be
\label{s}
s = \frac{|p_\phi|}{\alpha}\;.
\ee
The main hypothesis is the equivalence between the heavy-light block and the classical action of 5-line configuration
\be
f_{hl}\sim S_{cl}\;,
\ee
where $S_{cl}$ is the sum of the geodesic lengths with the particle masses being identified with the block scaled dimension parameters.\footnote{The equality is achieved with an appropriate identification between $q_{1,2}$ and the geodesic configuration boundary points.}

\subsection{Five-line configuration}
\begin{figure}[H]
\begin{center}
\begin{tikzpicture}[line width=1pt]\label{5pt}
\draw (0,0) circle (3cm);

\foreach \a in {1,2,...,40}{
\draw (\a*360/40: 3.5cm) coordinate(N\a){};
\draw (\a*360/40: 3cm) coordinate(K\a){};
\draw (\a*360/40: 2cm) coordinate(L\a){};
\draw (\a*360/40: 1.5cm) coordinate(I\a){};
\draw (\a*360/40: 1.1cm) coordinate(J\a){};
\draw (\a*360/40: 1cm) coordinate(M\a){};
;}


\draw plot [smooth, tension=1.0, line width=1pt] coordinates {(K35) (L34) (I30)};
\draw plot [smooth, tension=1.0, line width=1pt] coordinates {(K27) (L28) (I30)};
\draw plot [smooth, tension=1.0, line width=1pt] coordinates {(K23) (L23) (M25)};

\draw plot [smooth, tension=1.0, line width=1pt] coordinates {(M25) (0,0)};
\draw plot [smooth, tension=1.0, line width=1pt] coordinates {(I30) (J29) (M25)};


\fill  (I30) circle (0.8mm);
\fill  (M25) circle (0.8mm);

\fill  (0,0) circle (0.8mm);

\draw (1.6,-1.4) node {$1$};
\draw (-1.0,-1.9) node {$2$};
\draw (-1.9,-0.6) node {$3$};
\draw (-0.5,-0.1) node {$\tilde 2$};
\draw (-0.1,-0.8) node {$\tilde 1$};
\draw (0.2,-1.8) node {$A$};
\draw (-0.8,-1.0) node {$B$};
\draw (N23) node {$w_3$};
\draw (N27) node {$w_2$};
\draw (N35) node {$0$};

\end{tikzpicture}
\end{center}
\caption{Five-lines geodesic configuration at fixed time disk related to the 5-pt conformal block.}
\label{Config}
\end{figure}
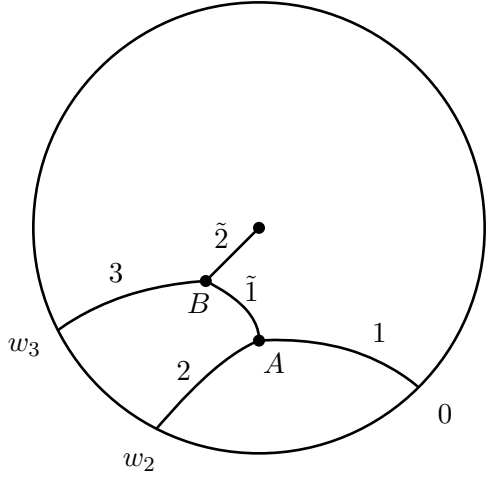

The goal of this section is to find the 5-particle geodesic length for the configuration depicted on the Fig. \bref{Config}
We denote by $S_{i}$ and $S_{\tilde i}$ the contribution of each particle to the total action
\be\label{fulllength}
S = \ep_1  S_{1}+\ep_2 S_{2}+\ep_3 S_{3}+\ept_1 S_{\1}+\ept_2 S_{\2}\;.
\ee
We minimize \eqref{fulllength} and obtain in particular:
\be
\label{equilibrium}
\big(\tilde\epsilon_1 \tilde p_\mu^1 + \epsilon_1 p_\mu^1+\epsilon_2 p_\mu^2\big)\;\Big|_{A}  = 0\;, \quad\big(\tilde\epsilon_1 \tilde p_\mu^1+\tilde \epsilon_2 \tilde p_\mu^2 +\epsilon_3 p_\mu^3 \big)\;\Big|_{B}  = 0\;,
\ee
arising as a result of the variation with respect to the vertexes positions.
The angular momenta conservation laws imply
\be
\label{ang}
\tilde s_2 = 0\;,
\qquad
\epsilon_3 s_3 - \tilde \epsilon_1 \tilde s_1 = 0\;,
\qquad
\epsilon_1 s_1 - \epsilon_2 s_2 - \tilde \epsilon_1 \tilde s_1 = 0\;,
\ee
where we suppose that proper time is growing with moving from the vertexes and it allows us to fix signs. We can resolve the radial equilibrium conditions with respect to the radial positions of the vertexes:
\be
\label{radA}
 \eta_A  =  \frac{1 - \sigma^2 }{s_1^2+s_2^2 -2 s_1s_2 \sigma  } \;,
\qquad\qquad
\sigma = \frac{\epsilon_1^2 +\epsilon_2^2 - \tilde\epsilon_1^2}{2\epsilon_1\epsilon_2}\;,
\ee
\be
\label{radB}
\eta_B  = -
\frac{(\epsilon_3 - \tilde \epsilon_1 - \tilde \epsilon_2) (\epsilon_3 + \tilde \epsilon_1 - \tilde \epsilon_1)
(\epsilon_3 - \tilde\epsilon_1 + \tilde\epsilon_2) (\epsilon_3 + \tilde\epsilon_1 + \tilde\epsilon_2)}{4\, \epsilon_3^2\,\tilde\epsilon_2^2\, s_3^2}\;.
\ee
The equations \eqref{ang},\eqref{radA},\eqref{radB} can be resolved for $s_1$ and $s_3$. We want to relate these two momenta with angular positions of two particles $w_2$ and $w_3$ (the first coordinate is fixed as reference point due to the axial symmetry). We can integrate angular coordinate along the path and express the final angular position through the integration constants and the coordinates of the vertexes:
\begin{eqnarray}
\label{firsteq}
e^{i\alpha w_2}=
\frac{\big(\sqrt{1-s_1^2\, \eta_A}-i s_1 \,\sqrt{1+\eta_A}\big)\big(\sqrt{1-s_2^2\, \eta_A}-i s_2\, \sqrt{1+\eta_A}\big)}
{(1-i s_1) (1-i s_2) }\;,
\end{eqnarray}
\be
\label{secondeq}
e^{i\alpha w_3}= \frac{\big(\sqrt{1-s_3^2 \eta_B }-i s_3 \sqrt{1+\eta_B}\big)
\big(\sqrt{1-\tilde s_1^2 \eta_B}-i \tilde s_1 \sqrt{1+\eta_B}\big)\big(\sqrt{1-s_1^2 \eta_A}-i s_1\sqrt{1+\eta_A}\big)}{(1-i s_3)\big(\sqrt{1-\tilde s_1^2 \eta_A}-i \tilde s_1\sqrt{1+\eta_A}\big)(1-i s_1)}\;.
\ee
The system of equations \eqref{ang}--\eqref{secondeq} completely defines the momenta and the radial coordinates of the vertexes as functions of the rescaled dimensions and two boundary coordinates. Partial lengths are following:
\be
\label{L1}
S_1 =  - \ln \frac{\sqrt{\eta_1}}{\sqrt{1+\eta_1}+\sqrt{1  - s_1^2\eta_1}} - \ln 2\Lambda\;,
\ee

\vspace{2mm}

\be
\label{L2}
S_2 =  - \ln \frac{\sqrt{\eta_1}}{\sqrt{1+\eta_1}+\sqrt{1- s_2^2\eta_1}} - \ln 2\Lambda \;,
\ee
\vspace{2mm}
\be
\label{L3}
S_{3} = - \ln \frac{\sqrt{\eta_2}}{\sqrt{1+\eta_2}+\sqrt{1- s_3^2\eta_2}} - \ln 2\Lambda\;,
\ee

\vspace{2mm}

\be
\label{La}
S_{\1} = \ln \frac{\sqrt{\eta_1}}{\sqrt{1+\eta_1}+\sqrt{1- (\tilde s_1 )^2 \eta_1}}- \ln \frac{\sqrt{\eta_2}}{\sqrt{1+\eta_2}+\sqrt{1- (\tilde s_1 )^2 \eta_2}} \;,
\ee
\vspace{2mm}
\be
\label{Lb}
S_{\tilde{2}} = \ln \frac{\sqrt{\eta_2}}{1+ \sqrt{1+\eta_2}}\;,
\ee
where we have introduced cutoff $\Lambda$.

\subsection{Perturbative solution}
In order to make final formulas less bulky, we set $\ep_1=\ep_2$. Similar consideration can be easily performed for unequal external dimensions. We start from one in some sense trivial solution of angular equations. Setting
\be
\begin{cases}\label{0gen}
   \quad \ep_3=0\;,
   \\
  \quad\ept_1=\ept_2\;,
 \end{cases}
\ee
one obtains
\be
\tilde s_1=\tilde s_2=0\;,
\ee
so that the four-point heavy-light block corresponds to the following action \cite{Hijano:2015qja}:
 \be
\label{Act1}
S(\ep_3=0,\ept_1=\ept_2) =  - 2 \epsilon_1 \ln\sin \theta_2 + \tilde\epsilon_1  \ln\tan\frac{\theta_2}{2}
 \;,
\ee
where $\theta_{2,3}=\frac{\alpha w_{2,3}}{2}$. In order to obtain the  5-pt block, we have to take into account an additional external field. Following \cite{Alkalaev:2015wia}, the  perturbed solution  can be found with  the assumptions:
\be
\begin{cases}\label{1gen}
   \quad \nu=\frac{\ep_3}{\ept_1}\ll 1\;,
   \\
  \quad\ept_1=\ept_2\;.
 \end{cases}
\ee
The calculation in this setting leads to the action, which corresponds to the 5-point heavy-light block with two equal intermediate dimensions:
\be
\label{Act2}
S(\ep_3\neq 0,\ept_1=\ept_2)  =  - 2 \epsilon_1 \ln\sin \theta_2 + \tilde\epsilon_1  \ln\tan\frac{\theta_2}{2}
- \epsilon_3 \ln \sin(2 \theta_3-\theta_2) +\mathcal{O}(\epsilon_3^2)\;.
\ee
We consider now the case, where $\ept_1\ne\ept_2$. Naively, such an assumption leads to the divergence of the radial coordinate of the B-vertex (see eq.\eqref{radB}).
However, this problem can be solved in the following way.
The character of the divergence implies that the difference between the two internal dimensions must be proportional to $\nu$. We introduce a new parameter $\mu$ in such a way that the previous case \eqref{1gen} is reproduced when this parameter is equal to zero, and vice versa, if $\nu=0$ we must obtain \eqref{0gen} because the internal dimensions are related, in fact, to one field. As a result, we can express the conditions in the following form:
\be
\begin{cases}\label{2gen}
   \quad \nu=\frac{\ep_3}{\ept_1}\ll 1\;,
   \\
  \quad\ept_2=\ept_1(1+\nu \mu)\;.
 \end{cases}
\ee
With these assumptions we obtain the following angular momenta
\begin{align}
&s_1=-\cot \theta_2-\nu\frac{\ept_1}{2\ep_1}\cot(2\theta_3-\theta_2) +\frac{\ept_1}{2\ep_1}\csc\theta_2+\mu\nu\frac{\ept_1}{2\ep_1}\csc(2\theta_3-\theta_2)\;, \\
&s_2=-\cot \theta_2+\nu\frac{\ept_1}{2\ep_1}\cot(2\theta_3-\theta_2) +\frac{\ept_1}{2\ep_1}\csc\theta_2-\mu\nu\frac{\ept_1}{2\ep_1}\csc(2\theta_3-\theta_2)\;, \\
&s_3=-\cot (2\theta_3-\theta_2)+\frac{\mu}{\sin(2\theta_3-\theta_2)}-\frac{\nu}{4}\frac{\left(3+\cos(4\theta_3-2\theta_2)-2\cos(2\theta_3-2\theta_2)-2\cos2\theta_3\right)}{\sin(2\theta_3-\theta_2)^3}\nonumber\\
&\quad\quad -\frac{\mu\nu}{4}\frac{\left(6\cos\theta_2+\cos(4\theta_3-\theta_2)+\cos(4\theta_3-3\theta_2)-8\cos(2\theta_3-\theta_2)\right)}{\sin(2\theta_3-\theta_2)^3}\;,\\
&\tilde s_1=-\cot(2\theta_3-\theta_2)+\mu\nu\csc(2\theta_3-\theta_2)\;,
\end{align}
and the following positions of vertexes:
\begin{align}
&\eta_A=\frac{4\ep_1-\ept_1^2}{(\ept_1\csc\theta_2-2\ep_1\cot\theta_2)^2}\;, \\
&\eta_B=\tan(2\theta_3-\theta_2)^2+2\mu\frac{\tan(2\theta_3-\theta_2)^2}{\cos(2\theta_3-\theta_2)}
-\nu \frac{1-2\cos\theta_2 \sec(2\theta_3-\theta_2)+\sec(2\theta_3-\theta_2)^3}{\cos(2\theta_3-\theta_2)}\nonumber\\
&\qquad+\mu\nu\left(1-3\sec(2\theta_3-\theta_2)^4+2\cos\theta_2\sec(2\theta_3-\theta_2)(2\sec(2\theta_3-\theta_2)^2-1)\right)\;.
\end{align}
Now we substitute the obtained momenta and the vertexes positions in the expression for the geodesic lengths:
\begin{align}
&S_1=-\log(\sin\theta_2)-\nu\cot(2\theta_3-\theta_2)\sin\theta_2+\mu\nu\frac{\sin\theta_2}{\sin(2\theta_3-\theta_2)}\;, \\
&S_2=-\log(\sin\theta_2)+\nu\cot(2\theta_3-\theta_2)\sin\theta_2-\mu\nu\frac{\sin\theta_2}{\sin(2\theta_3-\theta_2)}\;,\\
&S_3=-\log(-\sin(2\theta_3-\theta_2))+O(\nu)\;,\\
&S_{\tilde 1}=\log(\tan\frac{\theta_2}{2})-\log\frac{(2\ep_1\cos\theta_2-\ept_1)\sin(2\theta_3-\theta_2)}{(\cos\theta_2+1)(\mu-\cos(2\theta_3-\theta_2))}+\nu\frac{1}{4\sin(2\theta_3-\theta_2)^2}\times\nonumber\\
&\qquad\times\bigg(3+\cos(4\theta_3-2\theta_2)-4\cos\theta_2\cos(2\theta_3-\theta_2)+4\mu(\cos\theta_2-\cos(2\theta_3-\theta_2))\bigg)\;,\\
&S_{\tilde 2}=\log(\tan(\theta_3-\frac{\theta_2}{2})) -\nu\frac{1}{4\sin(2\theta_3-\theta_2)^2}\times\nonumber\\
&\qquad\times\bigg(3+\cos(4\theta_3-2\theta_2)-4\cos\theta_2\cos(2\theta_3-\theta_2)+4\mu(\cos\theta_2-\cos(2\theta_3-\theta_2))\bigg)\;,
\end{align}
We now have assembled all the ingredients that we need in order to write down the total action. Using
\be
\label{Map}
\theta_2=\frac{i \alpha}{2}\log(1-q_1q_2)\;,\quad\theta_3=\frac{i \alpha}{2}\log(1-q_2)
\ee
and expanding the total geodesic length  up to the first order in $\ep_3$, we get
\begin{multline}
\label{Action}
S=-2\ep_1\log\sinh\bigg(\frac{\alpha}{2}\log(1-q_1q_2)\bigg)+\ept_1\log\tanh\bigg(\frac{\alpha}{4}\log(1-q_1q_2)\bigg)\\
-\ep_3\log\sinh\bigg(\alpha\frac{2\log(1-q_2)-\log(1-q_1q_2)}{2}\bigg)+\ep_3\mu\log\tanh\bigg(\alpha\frac{2\log(1-q_2)-\log(1-q_1q_2)}{4}\bigg)\;.
\end{multline}
The corresponding result for the heavy-light block in the same approximation is
\begin{multline}
f_{hl}=-S-\ep_3\log\bigg(- \frac{\alpha q_2}{1-q_2}\bigg)+\ep_3\mu\log\bigg(-\frac{\alpha q_2}{2}\bigg)+\ept_1\log\bigg(-\frac{\alpha q_1 q_2}{4}\bigg)-2\ep_1\log\bigg(-\frac{\alpha q_1 q_2}{2}\bigg)\\
+\ep_1\log(1-q_1q_2)\;.
\end{multline}
The terms in the right hand side are connected with the standard conformal block prefactor and the prefactor, related to the conformal transformation of the fields \eqref{Map}.


\section{Conclusion}
We have considered the AdS/CFT correspondence between 5-point heavy-light classical conformal blocks
in the  boundary  CFT with a pure Virasoro algebra and classical worldline actions in the bulk.
We calculated the action for the 5-line geodesic configuration,
and after the conformal transformation \eqref{Map}, we obtained the equivalence between action \eqref{Action}
and the 5-point heavy-light block \eqref{f-hl-expan}--\eqref{g11} computed using $c$-recursion formulas \eqref{Recursion}.
We thus demonstrated that there exists a dual interpretation of the 5-point heavy-light block with unequal internal dimensions.

It is interesting to study possible generalizations of the holographic interpretation of conformal blocks
on a sphere and on higher-genus surfaces to the case of the boundary CFT with an extended symmetry algebra
and, in particular, to the case of W- and super-symmetry.
Another interesting question is to investigate the $1/c$ corrections for the conformal blocks in the context of the correspondence.

\vspace{10mm}
\noindent \textbf{Acknowledgements.}  The work of V.B. was performed with the financial support of the Russian Science Foundation (Grant No.14-50-00150). The work of R.G. has been funded by the  Russian Academic Excellence Project '5-100'. R.G. thanks ICTP-SAIFR and organizers of the Simons Non-perturbative Bootstrap School for warm hospitality. Authors are grateful to Konstantin Alkalaev for interesting discussions.

\newpage

\providecommand{\href}[2]{#2}\begingroup\raggedright\endgroup

\end{document}